\documentclass[mathleft
]{an}
\input{epsf}
\usepackage{graphicx}
\usepackage{times}
\def\la{\mathrel{\hbox{\rlap{\hbox{\lower4pt\hbox{$\sim$}}}{\raise2pt\hbox{$<$}}}}}
\def\ga{\mathrel{\hbox{\rlap{\hbox{\lower4pt\hbox{$\sim$}}}{\raise2pt\hbox{$>$}}}}}

\begin{document}

\Pagespan{789}{}
\Yearpublication{2010}%
\Yearsubmission{2010}%
\Month{TBC}%
\Volume{TBC}%
\Issue{TBC}%

\title{Unlocking the nature of ultraluminous X-ray sources using their X-ray spectra}

\author{Jeanette C. Gladstone\inst{1}\fnmsep\thanks{Corresponding author:
  \email{j.c.gladstone@ualberta.ca}\newline},
Timothy P. Roberts\inst{2} \and Chris Done\inst{2}
}
\titlerunning{Unlocking the nature of ULXs}
\authorrunning{J. Gladstone, T. Roberts, C. Done}
\institute{
Department of Physics, University of Alberta, Edmonton, Alberta, T6G 2G7, Canada
\and 
Department of Physics, University of Durham, South Road, Durham DH1 3LE, UK}

\received{TBC}
\accepted{TBC}
\publonline{TBC}


\abstract{We explore the nature of ultraluminous X-ray sources through detailed investigations of their spectral shape using some of the highest quality data available in the {\it XMM-Newton} public archives. Phenomenological models allow us to characterise their spectra, while more `physically-motivated' models enable us to explore the physical processes underlying these characteristics. 
These physical models imply the presence of extreme (probably super-Eddington) accretion on to stellar mass black holes. 
}
 
\maketitle

\section{Introduction}


For more than 30 years, bright ($L_X \ga 10^{39} $erg s$^{-1}$), extra-galactic, non-nuclear sources have been observed in X-rays, yet the nature of these ultraluminous X-ray sources (ULXs) is still unknown. Their luminosity precludes the possibility of standard stellar mass black holes (StMBHs) accreting isotropically below the Eddington limit, whilst their location within host galaxies - along with dynamical friction arguments - rules out super-massive black holes (SMBHs). The simplest explanation is that these objects are intermediate in both mass and luminosity between the black hole categories mentioned above, intermediate mass black holes (IMBH; Colbert \& Mushotzky 1999) that are accreting in a known state. An alternative is that they are StMBHs that are either exceeding the Eddington limit, by undergoing some form of super-Eddington accretion (e.g. Begelman 2002), or circumventing it via geometric or relativistic beaming (e.g. King et al. 2001; K{\"o}rding, Falcke \& Markoff 2002).

Simple phenomenological models revealed warm disc-like spectra in {\it ASCA} data, suggesting possible rapidly spinning (Kerr) StMBHs (Makishima et al. 2000). Increases in spectral quality allowed fitting of two component models (multi-coloured disc + power-law), indicating the presence of a soft excess at lower energies, well fit by a cool disc. The cool disc ($kT_{in} \sim 0.2$ keV) indicated the possible presence of an IMBH ($\sim 1000 M_\odot$, Miller et al. 2003). However, analysis of higher quality data showed evidence of a spectral break at a few keV (e.g. Stobbart, Roberts \& Wilms 2006, SRW06 hereafter), curvature not seen in standard states. Thus the accretion flows in ULXs may not simply be scaled up versions of those seen in Galactic black hole binaries.





\section{Characterising the spectra of ULXs}


Two main features have been highlighted by previous studies of ULX spectra: a soft excess, modelled by a cool disc with a power-law tail, and a break at higher energies. Gladstone, Roberts \& Done (2009; GRD09 hereafter) used some of the highest quality ULX data ($\ga$ 10,000 EPIC counts; based on SRW06) currently available in the {\it XMM-Newton} Science Archive (XSA\footnote{See {\tt http://xmm.esac.esa.int/xsa/}}) (a sample of 12 sources, with $L_X$ $\sim$ 10$^{39}$ -- a few 10$^{40}$ erg s$^{-1}$) to see if each of these features are present in the majority of ULX spectra. 

\begin{figure*}
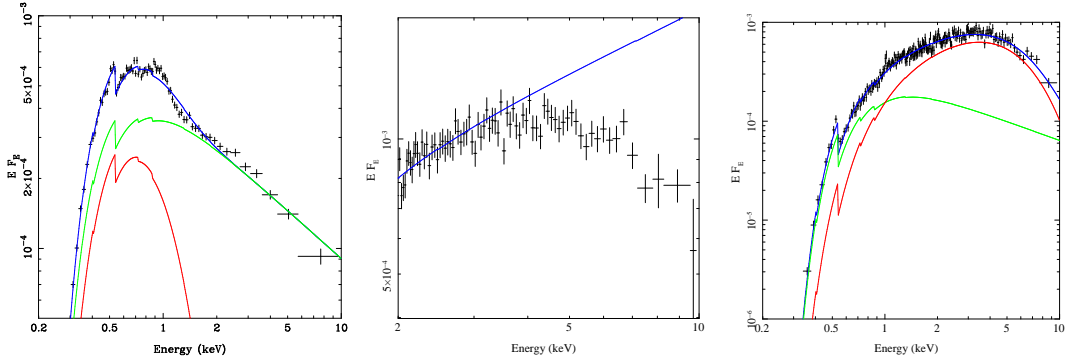

\leavevmode
\begin{center}
\includegraphics[height=47mm, angle=0]{figures/5408_imbh.ps} \hspace*{0.1cm}
\includegraphics[height=45mm, angle=0]{figures/1313_x2_pl.ps} \hspace*{0.1cm}
\includegraphics[height=47mm, angle=0]{figures/m81_x6_cool_disc.ps} 
\end{center}
\caption{{\it XMM-Newton} EPIC pn data for (a) NGC 5408 X-1, (b) NGC 1313 X-2 and (c) M81 X-6. In each case only pn data is shown and they have been rebinned for clarity. Disc components are plotted in red, power-laws are plotted in green and the combined model in blue. Plots (a) and (c) demonstrate the two types of fit received when combining an absorbed disc and power-law model. NGC 5408 X-1 (a)  has $kT_{in} \sim$ 0.19 kev, $\Gamma \sim$ 2.7, while M81 X-6 displays a warmer disc ($kT_{in} \sim$ 1.42 kev) with $\Gamma \sim$ 2.6. The first may indicate that the source is in the new {\it ultraluminous state} whilst the second spectrum is of a source described by a modified accretion disc (perhaps in the high, very high state). Plot (b) shows the clear presence of a break in the energy spectrum above 2 keV.}
\label{fig:pictures}
\end{figure*}

To look for a soft excess, each spectrum was fit with an absorbed power-law. A disc component (\textsc{diskbb} in \textsc{xspec}; Mitsuda et al. 1984) was then added to the model and $\Delta\chi^2$ found. An improvement ($\Delta \chi^2$ $>$ 30) was found in 11 spectra, an example of which can be seen in Figure \ref{fig:pictures}(a) (see also tables 4 \& 5 in GRD09 for results). Seven of the sample showed cool discs ($kT_{\rm in}<0.5$~keV), while the remaining sample displayed a more standard StMBH temperature. 

The possible break at higher energies was investigated by comparing power-law and broken power-law fits above 2~keV (see Figure \ref{fig:pictures}(b)). This provided clear evidence for a break in the energy spectrum above 3~keV in 11 of the 12 sources ($>$ 98 per cent statistical improvement using {\it F-test}; see also table 6 of GRD09). 

A soft excess and a break above $\sim$ 3 keV appear to be almost ubiquitous in this sample, although their spectra can be split into two distinct categories using these models. The first displays a warm disc and a break ($kT_{in} \ga$ 0.8 keV; $E_{break} \sim$ 4 keV). On closer inspection it is evident that each component is modelling the same spectral feature. In this case we find that the power-law contributes to emission at both higher and lower energies, broadening the disc model (see Figure \ref{fig:pictures}(c)). This is probably due to the \textsc{diskbb} providing a poor description of the broader spectrum expected from more realistic disc models (Done \& Davis 2008; Hui \& Krolik 2008). The second category exhibits a cool disc ($kT_{in} \la$ 0.8 keV) and break at higher energies (4 $\la E_{break} \la$ 7 keV), that are two distinct spectral features. This new combination of observational characteristics indicate that these systems are residing in an accretion state not commonly seen in Galactic X-ray binaries, an ``{\it ultraluminous state}''. 
To explore the nature of this new spectral state, we discuss more physically motivated models. 

\section{Testing physically motivated models}



\subsection{Slim discs}


The slim disc model suggests that the accretion disc can become so optically thick that energy released in the mid-plane does not have time to diffuse to the photosphere before being lost over the event horizon (Abramowicz et al. 1988). This causes the disc to swell, creating a geometrically `slim' disc. The change in disc structure causes a change in temperature gradient, represented by $r^p$. Some authors have looked for evidence of this changing disc structure with `{\it p-free}' (\textsc{diskpbb} in \textsc{xspec}; e.g. Watarai et al. 2001), searching for a shift from the standard value of $p$ ($\simeq 0.75$), to a value of $\sim$ 0.5 (e.g. Vierdayanti et al. 2006).

GRD09 applied this simplified version of the slim disc model to their sample. Initial results, at first glance, appeared to be in agreement with the slim disc model, with $p \sim$ 0.4 -- 0.6 (see GRD09 table 7). However, closer inspection revealed that while the inner-disc temperatures of some sources range from 1--3 keV, which can be expected for super-Eddington accretion, 4 of the sample exhibit higher temperatures (6 -- 13 keV), temperatures that are physically unrealistic. This appears to  indicate that the simplified slim disc model is an inadequate description of high quality ULX data. A more complex version of the slim disc (e.g. relativistic slim disc in Sadowski et al. 2010) should instead be applied to investigate this theory further.

\subsection{Comptonisation models}
\label{section:comp}

Comptonisation models have been applied to Galactic black hole binaries for many years with great success. GRD09 applied them to their sample as a starting point for further study. Two separate Comptonisation models were applied (\textsc{comptt}, Titarchuk 1994; \textsc{eqpair}, Coppi 1999), each in combination with a disc. 

The majority of the GRD09 sample were found to be parameterised by a cool accretion disc with a cool, optically thick corona ($\tau \ga$ 6), irrespective of Comptonising model. 
This is very different from standard black hole states. Only sources in the very high state (VHS) have parameters that are even close to those observed for ULXs (low state - $\tau \la$ 2, $kT_e \sim$ 50 keV; VHS - $\tau \sim$ 3 and $kT_e \sim$ 20 keV), and systems residing in this state are accreting at some of the highest known accretion rates. Here we observe greater optical depths and lower electron temperatures. This indicates that we may be observing ULX states that are more extreme than the VHS, and possibly super-Eddington accretion on to StMBHs. However, we must note that the discs remain cool, that could also be argued as support to IMBHs. 

There are intrinsic issues with the accretion structure described above. The model assumed that an optically thick corona does not intercept the line of sight to inner regions of the accretion disc, and that the underlying disc spectrum is independent of the presence of a corona (Kubota \& Done 2004). Each assumption is flawed. An optically thick corona could easily mask the inner regions of the disc, while two optically thick media in such close proximity requires that the energetics of the system must be considered. Both material and energy would be removed from the accretion disc in order to launch (and feed) the corona, depleting the accretion disc and altering its emission spectrum.

\begin{figure*}
\begin{center}
\leavevmode
\includegraphics[height=44mm, angle=0]{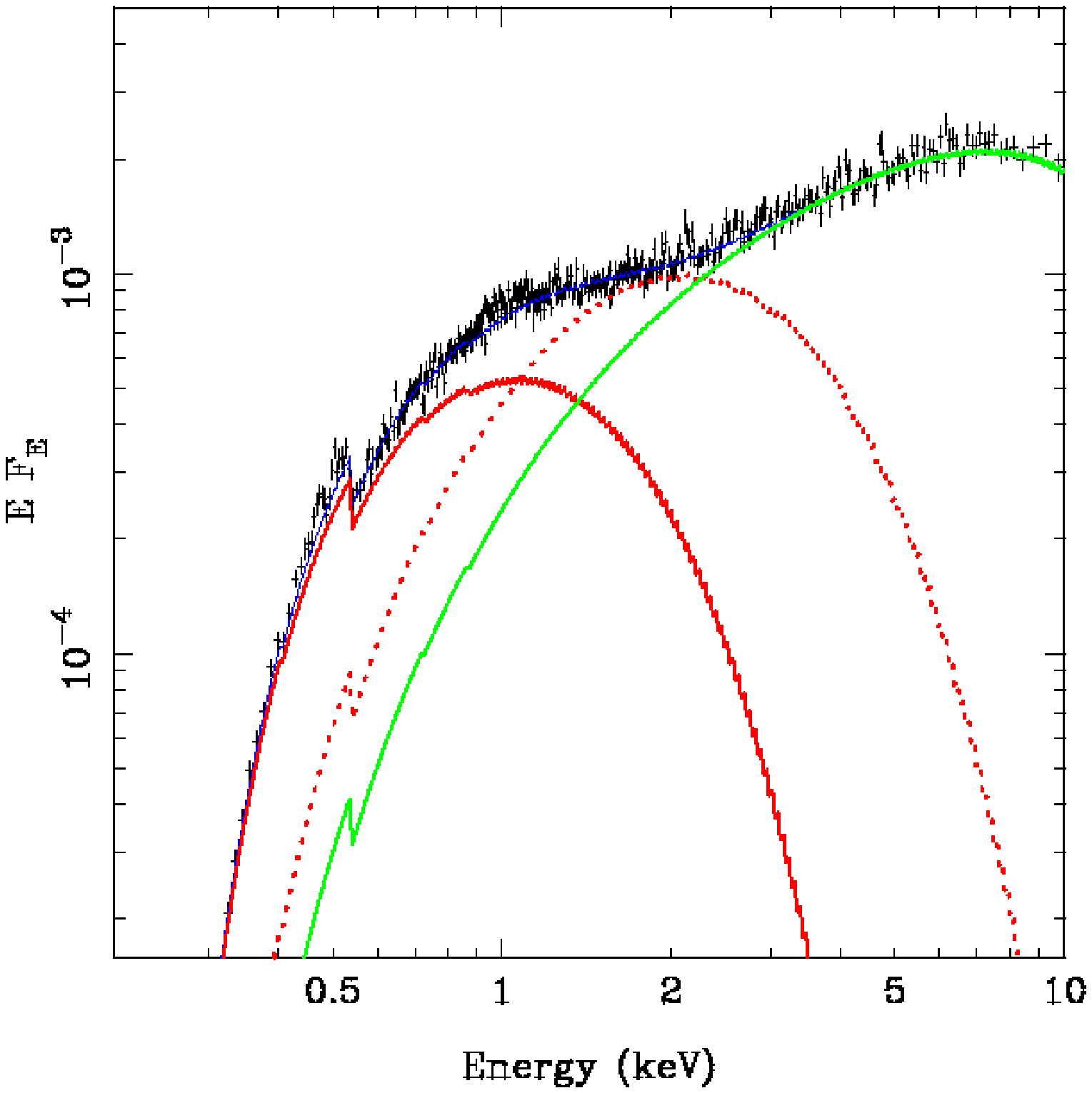} 
\includegraphics[height=44mm, angle=0]{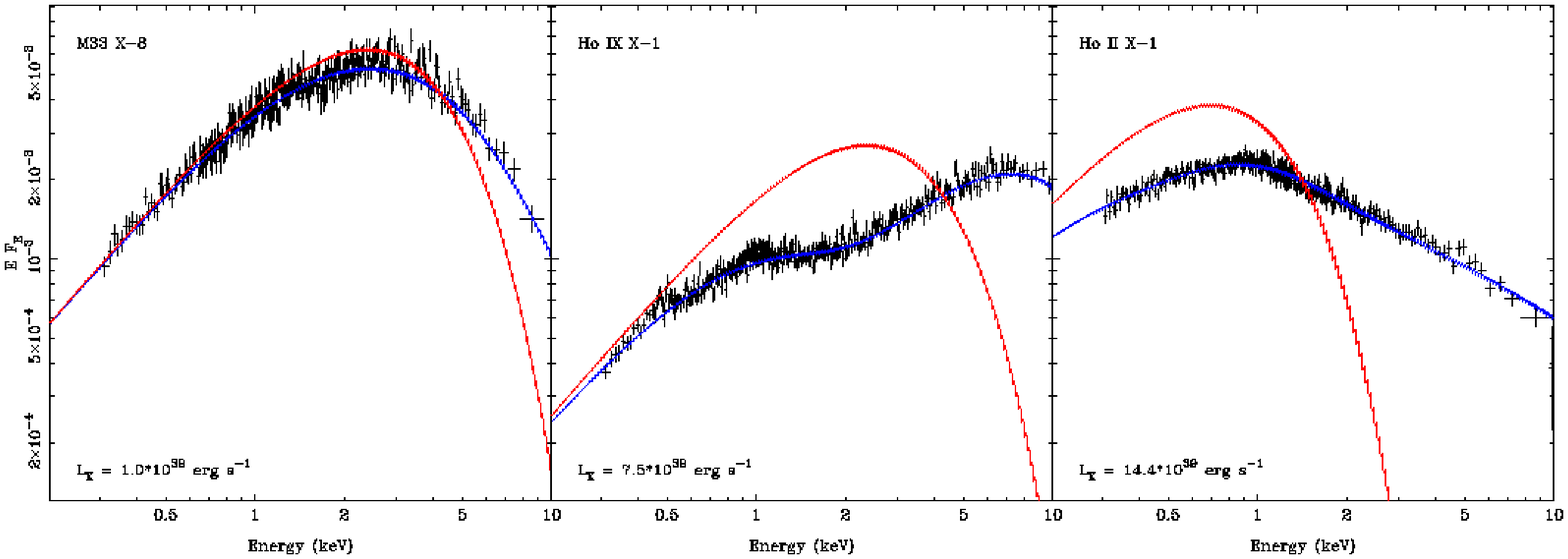} 
\end{center}
\caption{({\it Left}) {\it XMM-Newton} EPIC pn data from Ho IX X-1 (plotted in black and rebinned for clarity), fitted with an absorbed ultraluminous model (\textsc{dkbbfth}, in blue). Various components of the accretion system described by the model are overlaid. Disc components are shown in red whilst the corona is in green. The visible regions of the outer disc and the optically thick corona are plotted with a solid line, whilst the calculated masked emission from the cooled, energetically-coupled inner disc is plotted wited a dotted line. ({\it Right panels}) {\it XMM-Newton} EPIC-pn X-ray spectra for three ULXs from GRD09. The spectra are de-absorbed \& unfolded and show one spectrum from each suggested spectral regime: ($i$) the modified disc (M33 X-8); ($ii$) ultraluminous (Ho IX X-1); ($iii$) and extreme ultraluminous (possibly photosphere-dominated; Ho II X-1) states. The intrinsic disc spectrum, recovered when the corona is accounted for, is in red.}
\label{fig:chris}
\end{figure*}

\subsection{Energetically coupled disc and corona}
\label{section:chris}


In order to correct the flawed assumptions outlined above, GRD09 applied \textsc{dkbbfth} (Done \& Kubota 2006), incorporating the energetic disc-corona coupling model of Svensson \& Zdziarski (1994). It assumes that, as the accretion rate increases, material and energy are fed into the corona. By conserving energy, whilst calculating the fraction ($f$) dissipated in the corona, this model is able to retrieve details of the underlying disc including its resulting luminosity.

Figure \ref{fig:chris}({\it left}) shows Ho IX X-1, which clearly demonstrates that neither spectral feature gives direct information on the inner-disc temperature. The high energy turn over is described by the corona, whilst the soft excess provides the transitional radius (the outermost radius of the corona, beyond which the disc is clearly visible). The inner disc (dashed red line) is not visible to the observer. Figure 8 in GRD09 shows fits for the entire sample over-plotted with the recovered disc component. The recovered disc spectrum is the spectrum that would be observed if the corona (and its effects) could be removed from the system, as can be seen overlaid in three of the sample in Figure \ref{fig:chris}({\it right}). 

If we consider those systems initially characterised by a warm disc (plus power-law), we find that the recovered disc (from \textsc{dkbbfth}) is very similar to the actual spectrum (see Figure \ref{fig:chris}({\it right}-$i$)). These disc-like spectra are similar in shape to those of Galactic StMBHs residing in the high and very high state (when band-pass is considered; see also figure 9 in Done et al. 2007). This would indicate that they are possibly residing in the high or VHS. Of the remaining spectra, we find they two categories, those with a temperature that would indicate the presence of a StMBH (see Figure \ref{fig:chris}({\it right}-$ii$)), and those with a cool disc more indicative of an IMBH (see Figure \ref{fig:chris}({\it right}-$iii$)). The first group's parameters suggest that we are observing extreme StMBHs residing in the ultraluminous state, probably accreting at super-Eddington rates. In the second case, we may be observing IMBHs, or we may be observing something else. Theory predicts that a photosphere and wind are key components of super-Eddington accretion (e.g. Begelman et al. 2006; Poutanen et al. 2007) and this new spectral structure could provide observational evidence for such a photosphere. In this case, both mass and energy would be lost from the system whilst launching the wind. If this is the case, these sources are residing in an {\it extreme} ultraluminous state. In order to pursue this further, new models are required to fully explain the physics of such a system.

\subsection{Reflection models}


X-ray reflection models have also been proposed as a way to unlock the nature of ULXs, as X-ray reflection signatures have been noted in some StMBHs (e.g. Miller 2007) \& SMBHs (e.g. Tanaka et al. 1995). Caballero-Garc{\'i}a \& Fabian (2010) chose to apply this model to some of the highest quality data available in XSA. The authors proposed that ULX spectra could be explained by an X-ray irradiated disc around a rapidly spinning black hole. Here the authors used \textsc{reflionx} (Ross \& Fabian 2005) in combination with the Kerr kernel \textsc{kdblur} (Laor 1991) or \textsc{kerrconv} (Brenneman \& Reynolds 2006). The turn over at higher energies would be explained by a relativistically-broadened iron line, whilst an absorbed power-law models the underlying continuum, representing the emission source illuminating the disc. This implies that both the soft excess and the high energy break  are possible signatures of an X-ray irradiated accretion disc. The  authors also note that no statistical improvements are found with the addition of a disc component, they are therefore unable to comment on the mass of the black hole residing within ULXs, or on the implied accretion rate of these systems. 

The results of this study indicate that ULXs fall into two major groups.Some spectra appear to be dominated by a reflection component, whilst others are in a power-law dominated state. It is argued that these correspond to two of the regimes discussed in Miniutti \& Fabian (2004), in which the height of the primary source of X-rays is varied. This allows the observer to see differing amounts of the continuum and  reflection components.

\section{Spectral variability}


Recent work attempted the test the findings of Section \ref{section:chris} by studying a series of reasonable quality spectra of Holmberg IX X-1 (Vierdayanti et al. 2010a). {\it XMM-Newton} and {\it Swift} data (grouped by count rate) were combined to study spectral variations over long timescales, with the source luminosity varying by a factor of 3--4. When the data is fit with a combined disc and Comptonising corona model the results agree with those in Section \ref{section:comp}, the spectra are well fit by a cool optically thick corona. By comparing resultant fits the authors note that the temperature of the corona appeared to decrease, with a corresponding increase in optical depth, as the luminosity of the system increased. Pintore \& Zampieri (2010) also report on similar work  within these proceedings, results that appear to be in agreement with the predictions of super-Eddington accretion, and the ideas outlined by GRD09 (see the authors' conference proceedings for further details). However, the results of Vierdayanti et al. (2010a) also show further complexity in the spectral variability of these sources, with differences appearing in the spectral shape of these systems when the same level of flux at different times. In order to explore these differences in more detail, more in-depth studies are required into the spectral variability of these systems with high quality data. 

Another work published this year adds further support to GRD09, namely a study of the Galactic black hole binary GRS 1915+105, which has been observed to accrete at super-Eddington rates (given best estimates of mass and distance). The spectral evolution of this source shows similarities to that of ULXs, and the findings of  Section \ref{section:chris} (Vierdayanti et al. 2010b). When fitting extreme spectra of this source with a disc plus Comptonisation model, the authors found evidence of increasing optical depth with increasing luminosity. This provides strong support for the idea of extreme accretion onto StMBHs.

\section{The ultraluminous state?}

The ultraluminous state is an {\it observationally defined} state in which we observe the presence of a cool disc and a break in the spectrum above 3 keV. The questions that continue to surround this `state', and these sources, relate to the mass of the black hole residing at the centre of these systems and their accretion geometry. 

A range of models have been applied to the spectra of these systems in an attempt to answer these questions, with many suggesting extreme accretion onto StMBHs. Similar spectral evolution has been observed in the extreme Galactic StMBH GRS 1915+105, lending further support to this claim. 

At present, however, the results of such studies suggest that we need to exploit a wider bandpass to differentiate between these models. When comparing the fits from physically motivated models (excluding the slim disc scenario), we find that reduced $\chi^2$ values are very similar in almost all cases where there are overlaps in our samples. Therefore with the current data, it appears that an unambiguous determination of the accretion geometry of ULXs using single snapshot {\it XMM-Newton} observations alone is unlikely.  Walton et al. (2011) show that a wider bandpass can at least differentiate optically-thick corona models from reflection-dominated spectra. An alternative is to turn to different analysis methods, such as spectral or temporal variability with high quality data. The early results of such studies appear to support the predictions of the energetically coupled disc and corona model, but larger studies of such high quality X-ray data are required to confirm this. 




\end{document}